\documentclass[3p,times]{elsarticle}

\usepackage{ecrc}

\usepackage{epstopdf}

\volume{00}

\firstpage{1}

\journalname{Nuclear Physics A}

\runauth{K. Jung}

\jid{nupha}

\jnltitlelogo{Nuclear Physics A}

\usepackage{graphicx}
\usepackage{amsmath,amssymb}
\newcommand{\pt}{$\rm{p}_{\rm T}$}
\begin{document}

\begin{frontmatter}



\title{Measurements of b-jet Nuclear Modification Factors in pPb and PbPb
  Collisions with CMS}

%% Single author (and collaboration) - please insert
\author{Kurt Jung (for the CMS\fnref{col1} Collaboration)}
\fntext[col1] {A list of members of the CMS Collaboration and acknowledgements can be found at the end of this issue.}
\address{Department of Physics and Astronomy, Purdue University, 525
    Northwestern Ave., West Lafayette, IN, USA}

%% For multiple authors, replace the above by:

%\author[label1]{Author1}
%\author[label2]{Author2}

%\address[label1]{Address 1}
%\address[label2]{Address 2}

\begin{abstract}
We present measurements of the nuclear modification factors $R_{AA}$
and $R_{pA}^{PYTHIA}$ of b jets in lead-lead and proton-lead
collisions, respectively, using the CMS detector.  Jets from b-quark
fragmentations are found by exploiting the long lifetime of the
b-quark through tagging methods using distributions of the secondary
vertex displacement.  From these, b-jet cross-sections are calculated
and compared to the pp cross-section from the 2.76 TeV pp data
collected in 2013 and to a PYTHIA simulation at 5.02 TeV, where these
center-of-mass energies correspond to those of the PbPb and pPb data.
We observe significant suppression for b jets in PbPb, and a
$R_{pA}^{PYTHIA}$ value
consistent with unity for b jets in pPb.  Results from both collision species show
remarkable correspondance with inclusive-jet suppression measurements,
indicating that mass-dependent energy-loss effects are negligible at
\pt\ values greater than around 50 GeV/c.  We use 150 $\rm{\mu}$b$^{-1}$ of
lead-lead data and 35 nb$^{-1}$ of proton-lead data collected at the LHC.
\end{abstract}


\end{frontmatter}

%\linenumbers


\section{Introduction}
\label{intro}
Quenching of jets in heavy-ion collisions is expected to depend
heavily on the mass of the fragmenting parton.  Under the assumption
that gluon radiation is the dominant energy-loss mechanism, jets from heavy
quarks are expected to radiate less due to the ``dead-cone effect'',
especially when the parton \pt\ is comparible to the parton mass.
It must be said, however, that the mechanisms for in-medium partonic energy-loss
are still poorly constrained.  These measurements of the energy
loss observed in jets from heavy-ion collisions as a function of jet flavor provide
powerful constraints on the understanding of possible energy-loss
mechanisms, as jet flavor is a direct proxy for the different parton
masses.  This analysis will focus on b-jet energy loss.

CMS is described in detail in the original detector publication
\cite{cmsexp}, and its silicon tracker and hadronic calorimeter
are excellent experimental tools for observing heavy flavor jets in heavy-ion
collisons.  Jets formed from heavy
flavor quark fragmentation are typically tagged in one of two ways:
first, by the direct reconstruction of a displaced vertex, and second by
the displacement of individual tracks.  Using the track-only tagging
method as a cross-check ensures the secondary
vertex reconstruction selections remain unbiased.  The
three-dimensional distance of the closest track point to the primary
vertex is defined as the impact parameter \cite{BTV-12-001}. 
Information from these tracks
and vertices are typically combined into a quantity which optimizes
their discrimination between heavy and light flavor jets.  In this
analysis, we use a discriminator to tag b jets which is based on the
displacement of the reconstructed secondary vertex (SV) with
respect to the primary vertex of the interaction.  This discriminator
is called the Simple Secondary Vertex tagger (SSV) \cite{BTV-12-001},
and is based on the displacement significance (displacement divided by its uncertainty) 
of reconstructed secondary vertices.  Detailed studies of quantities
related to secondary vertex reconstruction show that the SSV tagger is well-modeled by both PbPb
and pPb simulations, which means that it ought to perform well
in both collision species.  The efficiency of
this SV tagging is then evaluated both directly from simulation, and
with a data-driven method using a discriminator derived from an
impact parameter-based method.  This impact parameter tagger is only weakly
correlated to the secondary vertex tagger and therefore provides a reliable way
to evaluate the efficiency and purity of the SV tagger directly from
data.  Another discriminating variable, called Jet Probability (JP),
takes advantage of this impact parameter tagger.  The JP algorithm orders jet-associated
tracks based on their impact parameter significance and calculates the likelihood that they
come from the primary vertex \cite{BTV-12-001}.  The less
likely that tracks originate from the primary vertex, the more likely
they stem from a long-lived jet seed.  The advantage to the JP algorithm is that it
provides a measure of discrimination for nearly all b jets, even in
the case when no secondary vertex is reconstructed.   This property of the JP tagger
is exploited to obtain a data-driven estimate of the SSV tagging
efficiency \cite{BTV-12-001}.

The performance of lifetime-based tagging relies on the high
efficiency and low fake rate of reconstruction of charged particle
tracks from displaced vertices. In proton-lead collisions, the event
multiplicity is low enough so that every track can be iteratively
reconstructed, but this is not the case in lead-lead collisions.  
Due to timing and memory constraints, the standard heavy-ion tracking
algorithm \cite{hin10005} in CMS is largely restricted to the
reconstruction of charged particles from the primary vertex.
Therefore, to enhance the efficiency of track reconstruction from secondary
vertices, an additional reconstruction method is used.  This
``regional tracking'' algorithm uses 
reconstructed jets as seeds and limits the search
window for tracker hits to a region defined around the jet axis.
Once reconstructed, the tagging performance is typically benchmarked by
finding the b-jet tagging efficiency as a function of the light (udsg) jet and
charm jet misidentification rate, such that the resulting curves are
independent of the underlying b-jet fraction.
The performance of the b-tagging degrades somewhat with the increased
multiplicity of PbPb collisions, giving roughly a factor of three poorer
rejection of light jets for a b jet efficiency of 50\%, relative to a
{\textsc{PYTHIA}} simulation alone.  Despite the reduced performance, one is
still able to achieve roughly a factor of 100 rejection of light jets
for a b jet efficiency around 50\%.  The charm rejection for this
b jet efficiency is about a factor of 10.

%\begin{figure}[!b]
%\begin{center}
%\resizebox{0.49\textwidth}{!}{\includegraphics{./figures/bVsX_Both.pdf}}
%\end{center}
%\caption[b jet tagging efficiency vs. light jet mis-tag
%efficiency]{The b jet tagging efficiency vs. the charm and light jet
 % mis-tag efficiency for simulated pp events from {\textsc{Pythia}} (green) and simulated PbPb events from {\textsc{Pythia}} embedded in {\textsc{Hydjet}} (blue) for the SSVH%E discriminator.  The red cross marks the working point of the SSVHE discriminator used in this analysis.}
%\label{fig:bVsL}
%\end{figure}

The b-jet yield in each \pt\ bin is obtained via equation 1, where $f_{b}$ denotes the purity of the sample after b-tagging,
$\epsilon_{b}$ refers to the efficiency of the b jet tagger, and
$N_{all}$ is the total number of jets in the sample.

\begin{equation}
N_{b} = N_{all} \frac{f_{b}}{\epsilon_{b}}
\end{equation}

These purity and efficiency quantities are obtained by
fitting to MC templates and cross-checked against the b-jet purity and efficiency
found by using the JP tagger.  For the purity, distributions of the secondary
vertex mass in MC are fit to those obtained in data, where the fractional
contribution of each jet flavor is allowed to float.
%Fig. ~\ref{fig:svm100} shows an example of this fitting in a \pt\ bin of
%80 $<$ \pt\ $<$ 100 GeV/c, after applying the b-tagger. 
Contributions
from all three flavor types are comparable, but above about 2
GeV/c$^{2}$ (which roughly corresponds to the charm quark mass),
the jet sample is dominated by b jets.

%\begin{figure}[!t]
%\begin{center}
%\resizebox{0.49\textwidth}{!}{\includegraphics{./figures/fitPbPb_SVmass_80_100.pdf}}
%\resizebox{0.49\textwidth}{!}{\includegraphics{./figures/fitpp_SVmass_80_100.pdf}}
%\end{center}
%\caption{Template fits to the SV mass distributions in PbPb (left) and
 % pp (right) collisions, after tagging with the SSVHE discriminator
  %for jets of 80 $<$ \pt\ $<$ 100 GeV/c}
%\label{fig:svm100}
%\end{figure}

The efficiency ($\epsilon_{b}$) is obtained via simulation, and is
  cross-checked using the JP tagger, as described above.
  Distributions of the JP tagger discriminator are obtained both
  before and after applying the SSV-tagged selection, as a function of
  jet \pt.  
%Since the impact
 % parameter distributions can be calibrated in a data-driven way using
  %the negative values of the impact parameter distribution, we can
  %expect this method provides an accurate determination of the
  %efficiency in data \cite{CMS-PAS-BTV-11-004}.  
In practice, we find the difference between the
  data-driven and MC $\epsilon_{b}$ value is about 5\%.  This
  difference is taken as a systematic uncertainty.

\begin{figure}[!t]
\begin{center}
\resizebox{0.34\textwidth}{!}{\includegraphics{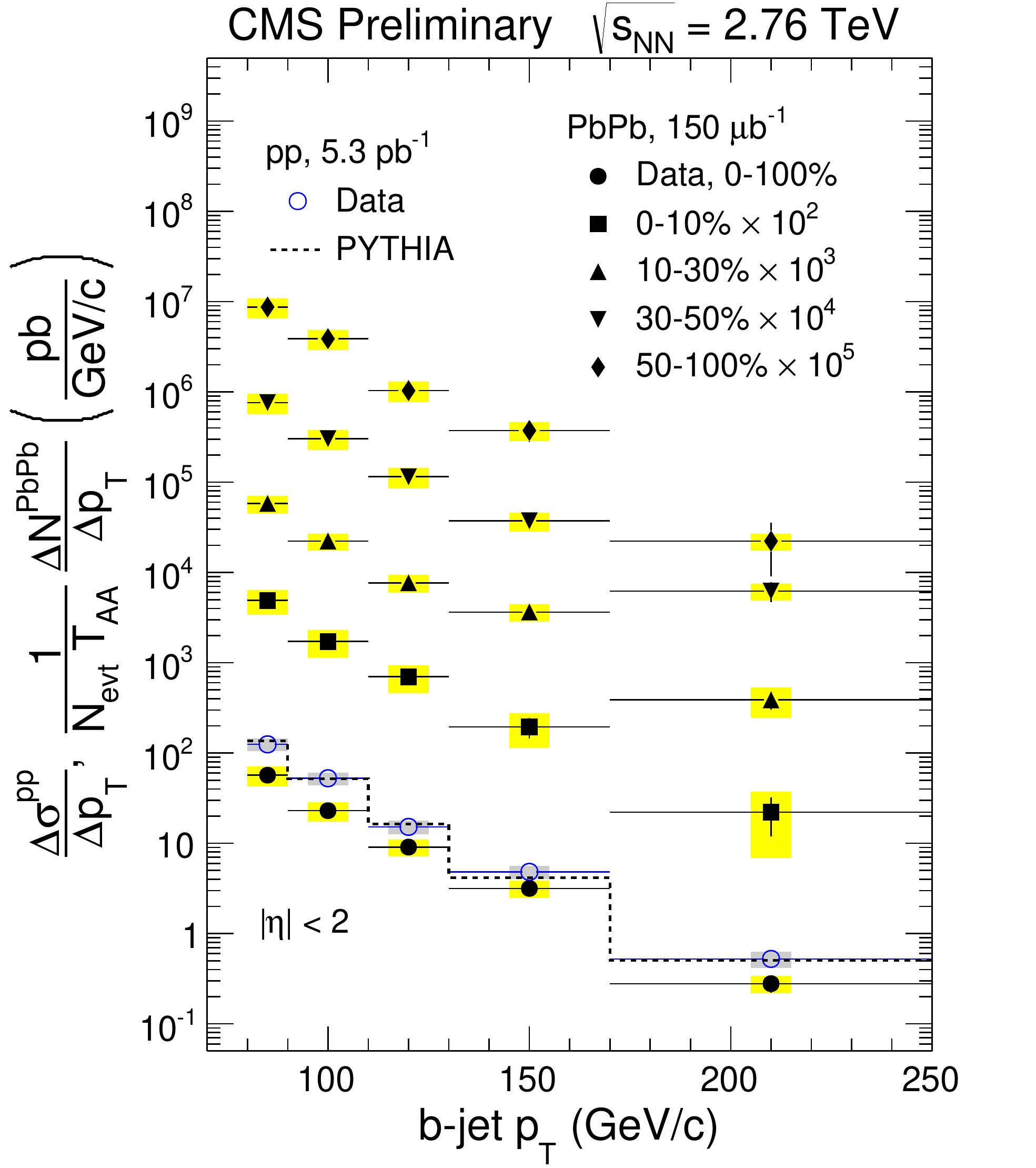}}
\resizebox{0.34\textwidth}{!}{\includegraphics{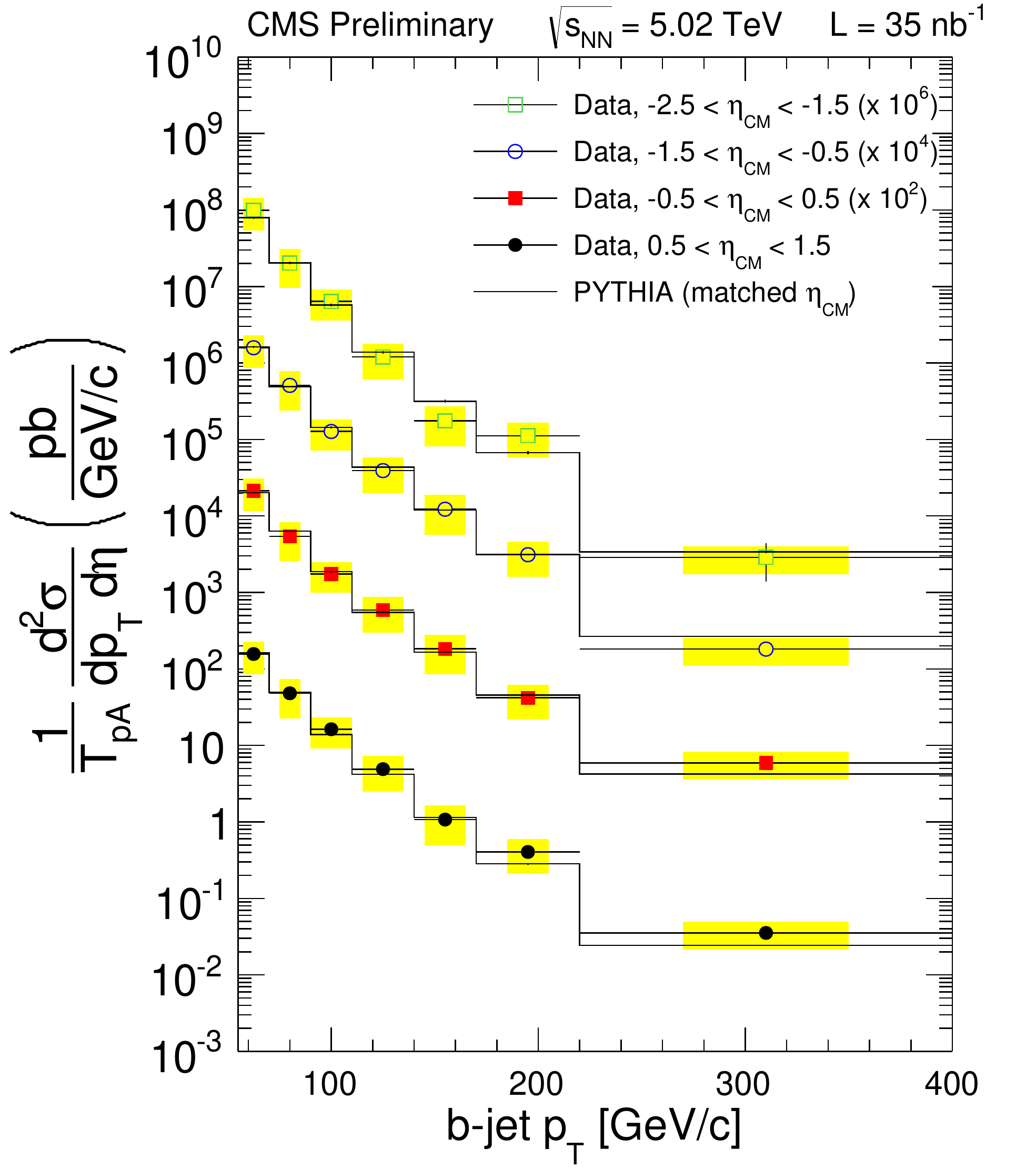}}
\end{center}
\caption{b-jet spectra as a function of jet \pt\ in bins of centrality
  for PbPb (left), and in bins of pseudorapidity for pPb (right)}
\label{fig:spectra}
\end{figure}

Finally, the b-jet yield is unfolded via the D'Agostini iterative
procedure \cite{dagostini:aa}, as implemented by the RooFit statistical modeling toolkit.  Once unfolded,
the PbPb (pPb) data is scaled by a Glauber scaling factor ``$T_{AA}$''
(``$T_{pA}$''), in order to compare directly to pp data (simulation) \cite{glauber}.  The values of
$T_{AA}$ and $T_{pA}$ are the number of nucleon-nucleon collisions divided by the
total inelastic cross-section, and can be interpreted as the
pp-equivalent luminosity per heavy-ion collision.  These b-jet cross-sections are
shown in figure \ref{fig:spectra} for various centrality selections in
PbPb (left) and for various pseudorapidity selections in pPb (right).
Also shown is the measured b jet cross-section in pp collisions as
well as simulations from the {\textsc{PYTHIA}} Z2 tune \cite{field:aa}, which agree well with the data.
The pp luminosity measurement has an uncertainty of about 3.6\%, while
the uncertainty in $T_{AA}$ varies from 4\% to 15\% as a function of
centrality.

\begin{figure}[!t]
\begin{center}
\resizebox{0.35\textwidth}{!}{\includegraphics{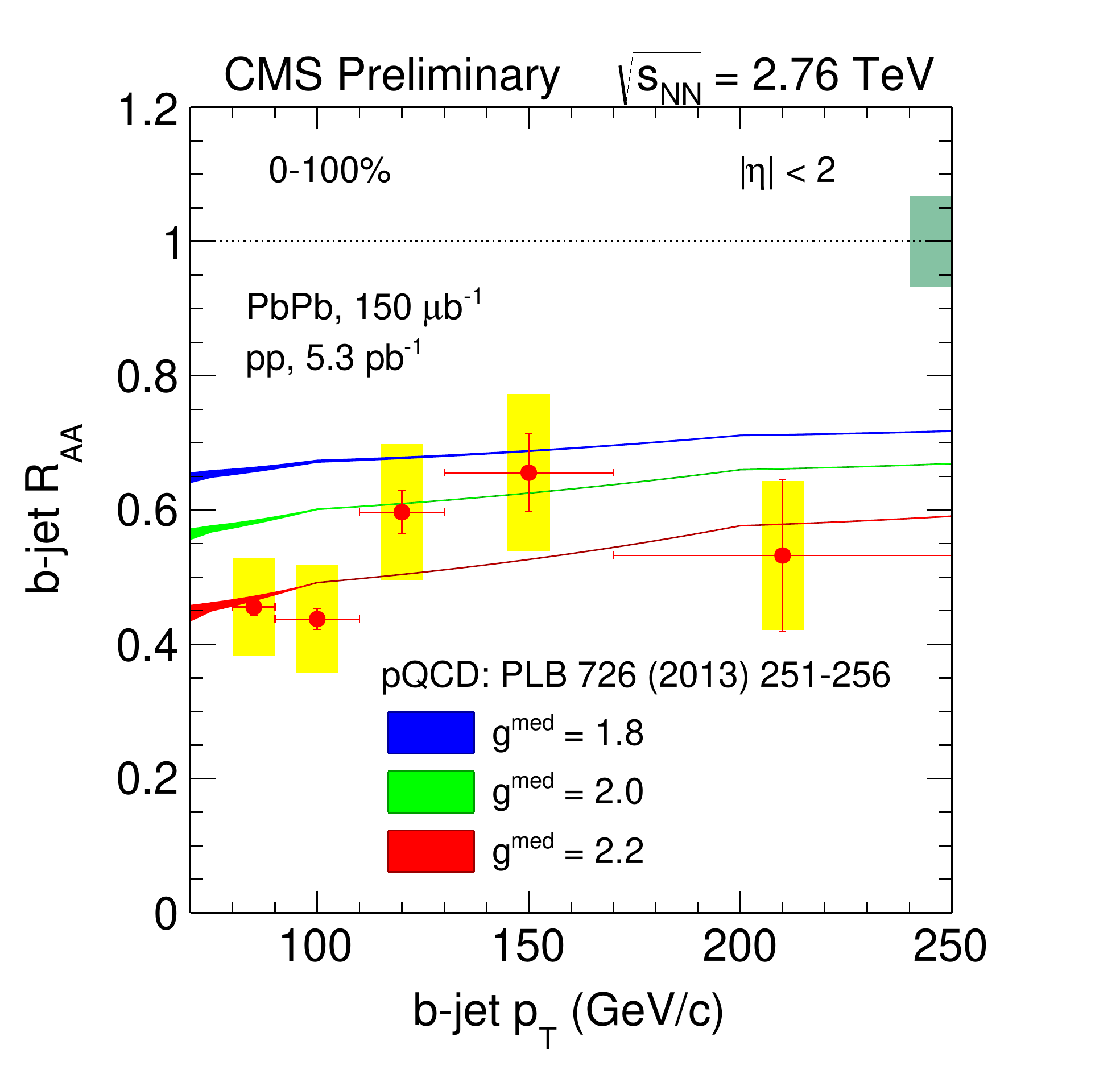}}
\resizebox{0.35\textwidth}{!}{\includegraphics{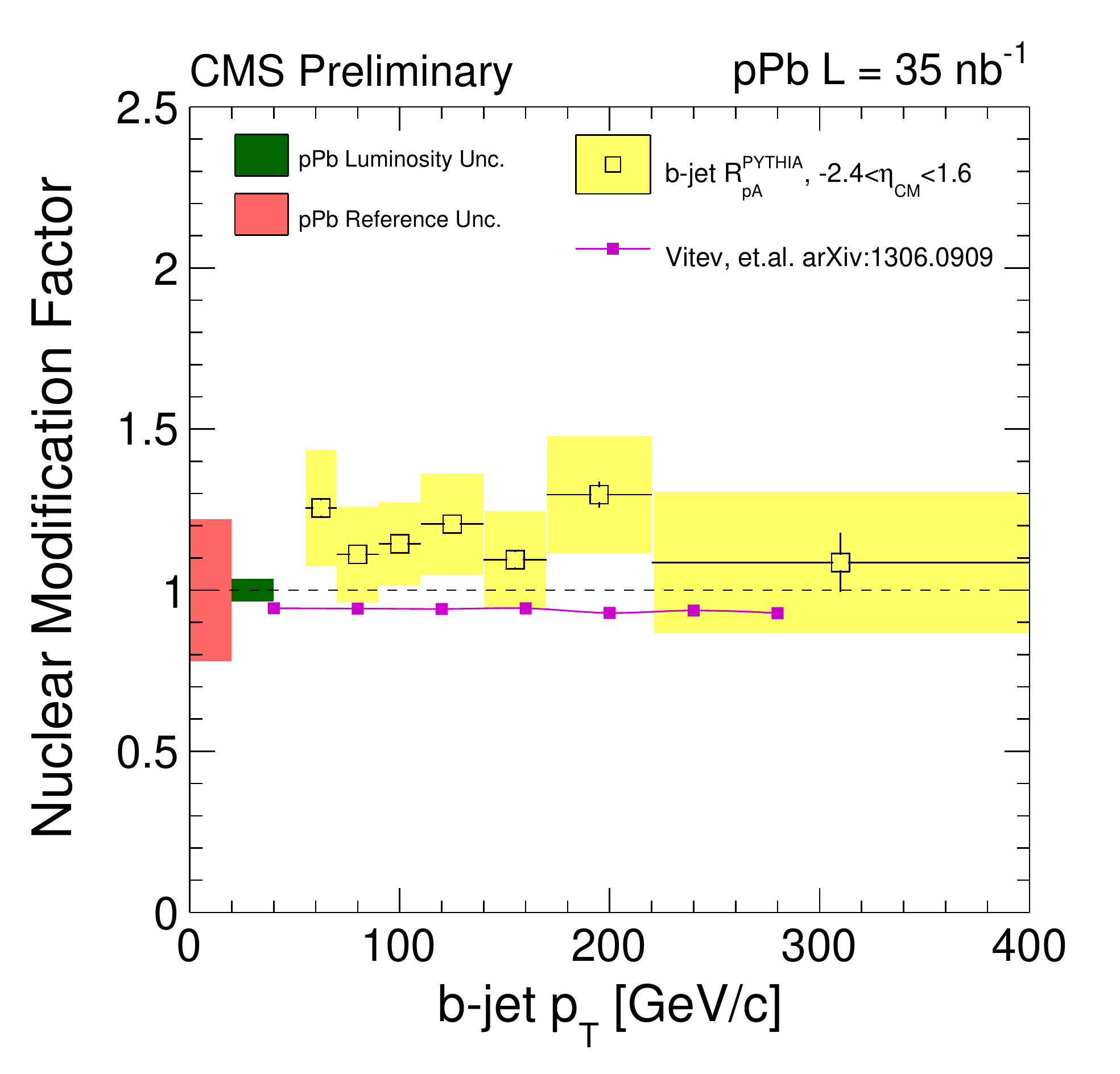}}
\end{center}
\caption{Centrality integrated b-jet $R_{AA}$ as a function of \pt\
  (left) and $R_{pA}^{PYTHIA}$ (right).
  The filled green box denotes the normalization uncertainty from the
  integrated luminosity in pp and the $T_{AA}$ in PbPb, while the
  filled red box denotes the uncertainty from the pp PYTHIA simulation
  used in the $R_{pA}$ calculation. Theoretical predictions are from
  \cite{vitev:aa}.}
\label{fig:bJetRpaRAA}
\end{figure}

Figure \ref{fig:bJetRpaRAA} shows the measurements of b-jet $R_{AA}$
(left) and $R_{pA}^{PYTHIA}$ (right).  A comparison of these two plots
shows that the suppression effects for the PbPb plot are significantly
greater than those observed in pPb, which indicates that initial state
effects are not responsible for the suppression.  In fact, as all
$R_{pA}^{PYTHIA}$ values are greater than one, indications
of a small enhancement may be drawn.  It is likely that any
enhancement seen in pPb collisions stems from the Cronin effect, where
multiple scatterings lead to an enhanced jet production from that
observed in pp collisions.  This also seems to be the case when a
comparison is made to a theoretical predicition from Vitev et. al. \cite{vitev:aa},
which is a calculation made in a perturbative QCD framework and an
identical jet reconstruction algorithm.  This prediction is shown as
the purple curve just below unity in Fig. \ref{fig:bJetRpaRAA}.

\begin{figure}[!t]
\begin{center}
\resizebox{0.35\textwidth}{!}{\includegraphics{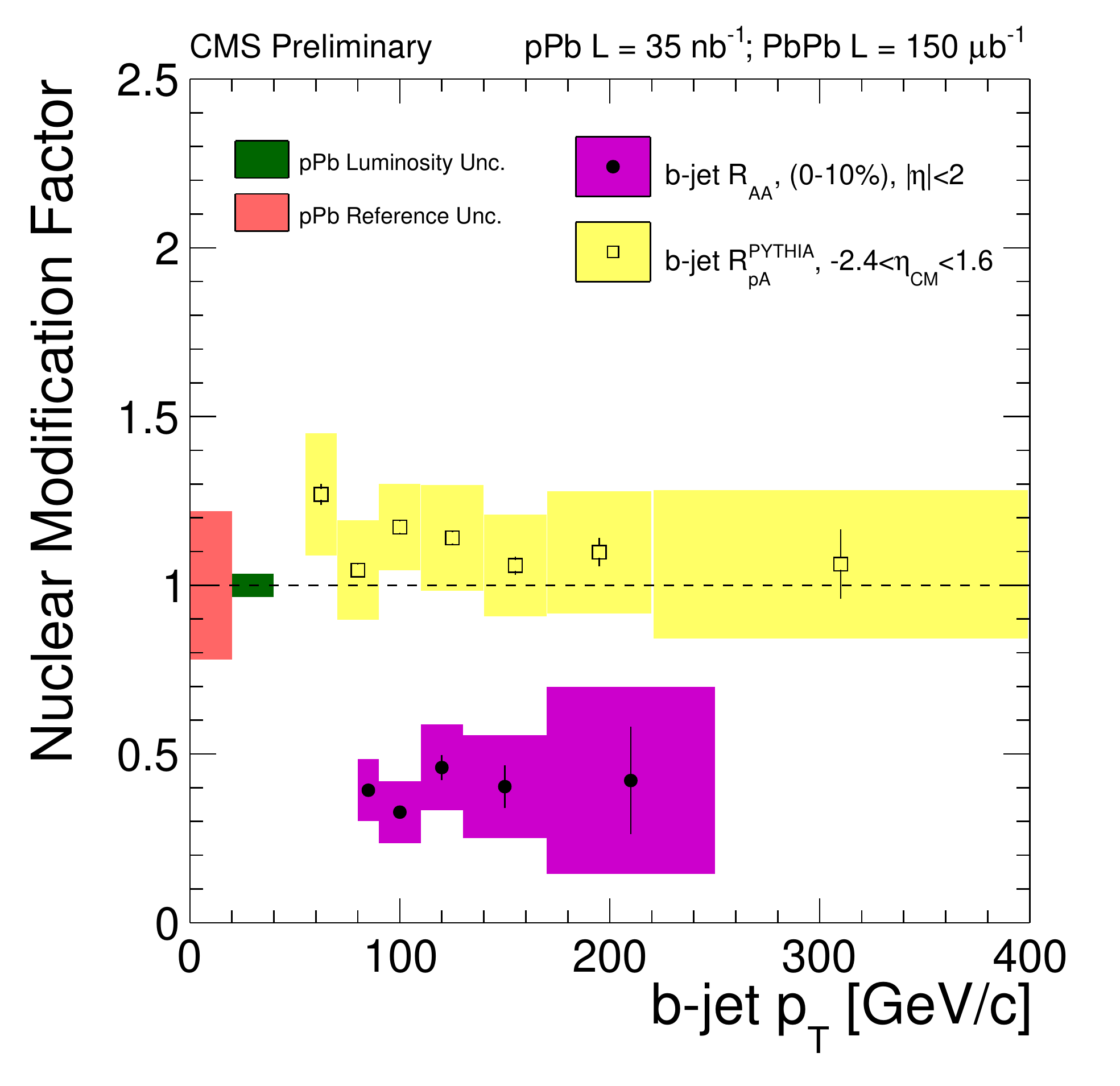}}
\resizebox{0.35\textwidth}{!}{\includegraphics{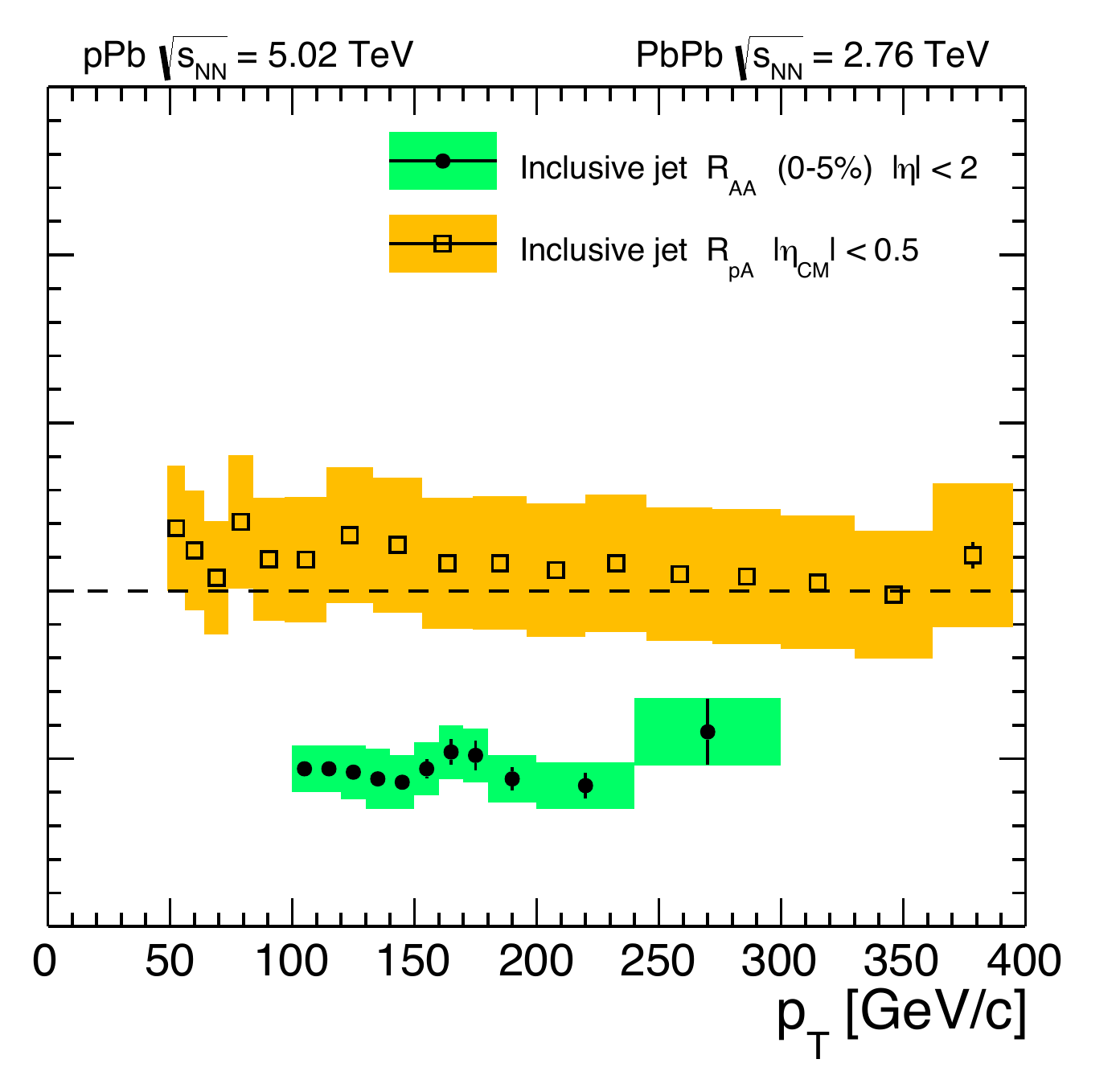}}
\end{center}
\caption{Nuclear modification factor comparison for b-jet
  $R_{pA}^{PYTHIA}$ and $R_{AA}$ (left) and
  inclusive-jet $R_{pA}^{PYTHIA}$ and $R_{AA}$ (right).  The $R_{AA}$
  values shown are for 0-10\% centrality, while the $R_{pA}$ values
  are for inclusive-centrality.  Note the similarity between both the
  $R_{pA}$ and $R_{AA}$ curves between the b-jet and inclusive-jet data.}
\label{fig:bJetvsInclJet_all}
\end{figure}

Finally, Fig. \ref{fig:bJetvsInclJet_all} shows measurements of both
$R_{pA}^{PYTHIA}$ and $R_{AA}$ as
a function of \pt\ for both the b-jets \cite{HIN-14-007, HIN-12-003} and the inclusive jets
\cite{HIN-14-001, HIN-12-004}, as denoted in the
legends.  We observe a very consistent result between the flavored
jets and the inclusive-jet measurements, indicating that the
suppression effects in PbPb collisions and enhancement effects in pPb
are roughly parton-mass independent at very high \pt.  

\section{Conclusions}
B jets in PbPb are found to be suppressed over a wide range of \pt, from
80-250 GeV/c.  Futhermore, the $R_{AA}$ value is found to decrease
with increasing collision centrality.  The b jets observed in pPb
show virtually no suppression effects and may show possible hints of a
Cronin enhancement due to cold nuclear matter effects.  This lack of
suppression is consistent within uncertainties across a wide range of
\pt\ and pseudorapidity.  Measurements as a function of pseudorapidity
may have implications regarding the
suppression effects as a function of the Bjorken-x value of the
incident parton in the lead nucleus (see \cite{HIN-13-001} for further details).

In the end, the data presented in this analysis shows two striking features.
First, the b jet production is clearly suppressed in PbPb collisions,
while the production in pPb collisions is consistent with that
calculated in pp simulations.  Second, the data shows that the
respective b-jet nuclear modification factors of both PbPb and pPb are
very consistent with those seen in the inclusive-jet studies from
CMS.  From these observations, we can draw two conclusions: first that
the suppression effects observed in PbPb collisions are not part of
initial state effects, since the pPb nuclear modification factor is
strikingly different than that observed in PbPb.  Second, the jet
suppression effects are essentially mass independent at very high
\pt.  These effects favor a perturbative energy-loss model where
these mass-dependent effects are expected to be small, and are in
contrast to an AdS/CFT inspired model where such effects can be quite
large, even at very high values of \pt\ \cite{Horowitz:2007su}.



%% References


\bibliographystyle{elsarticle-num}
\bibliography{QM14-template.bib}



%\begin{thebibliography}{00}


%\bibitem{ref1} J. van der Geer, J.A.J. Hanraads, R.A. Lupton, J. Sci. Commun. 163 (2000) 51Ð59. 
%\bibitem{ref2} W. Strunk Jr., E.B. White, The Elements of Style, third ed., Macmillan, New York, 1979. 

%\end{thebibliography}

\end{document}